\documentclass{sigchi}

\usepackage{balance}       % to better equalize the last page
\usepackage{graphics}      % for EPS, load graphicx instead 
\usepackage[T1]{fontenc}   % for umlauts and other diaeresis
\usepackage{txfonts}
\usepackage{mathptmx}
\usepackage[pdflang={en-US},pdftex]{hyperref}
\usepackage{color}
\usepackage{booktabs}
\usepackage{textcomp}
\usepackage{etoolbox}

\usepackage{microtype}        % Improved Tracking and Kerning
\usepackage{ccicons}          % Cite your images correctly!
\def\plaintitle{Adaptive App Design by Detecting Handedness}

\def\emptyauthor{}
\def\plainkeywords{One-handed design; UX design guidelines; mobile; grip detection;}

\makeatletter
\def\@copyrightspace{\relax}
\def\url@leostyle{%
  \@ifundefined{selectfont}{
    \def\UrlFont{\sf}
  }{
    \def\UrlFont{\small\bf\ttfamily}
  }}
\makeatother
\def\pprw{8.5in}
\def\pprh{11in}

\definecolor{linkColor}{RGB}{6,125,233}
\hypersetup{%
  pdftitle={\plaintitle},
  pdfauthor={\emptyauthor},
  pdfkeywords={\plainkeywords},
  pdfdisplaydoctitle=true, % For Accessibility
  bookmarksnumbered,
  pdfstartview={FitH},
  colorlinks,
  citecolor=black,
  filecolor=black,
  linkcolor=black,
  urlcolor=linkColor,
  breaklinks=true,
  hypertexnames=false
}

\usepackage{cuted}
\usepackage{capt-of}
\begin{document}

\title{\plaintitle}

\numberofauthors{2}
\author{%
  \alignauthor{Kriti Nelavelli\\
    \affaddr{Georgia Institute of Technology}\\
    \affaddr{Atlanta, GA USA}\\
    \email{kritin@gatech.edu}}\\
  \alignauthor{Thomas Pl\"{o}tz\\
    \affaddr{Georgia Institute of Technology}\\
    \affaddr{Atlanta, GA USA}\\
    \email{thomas.ploetz@gatech.edu}}\\
}

\maketitle
\begin{strip}\centering
\includegraphics[width=\textwidth]{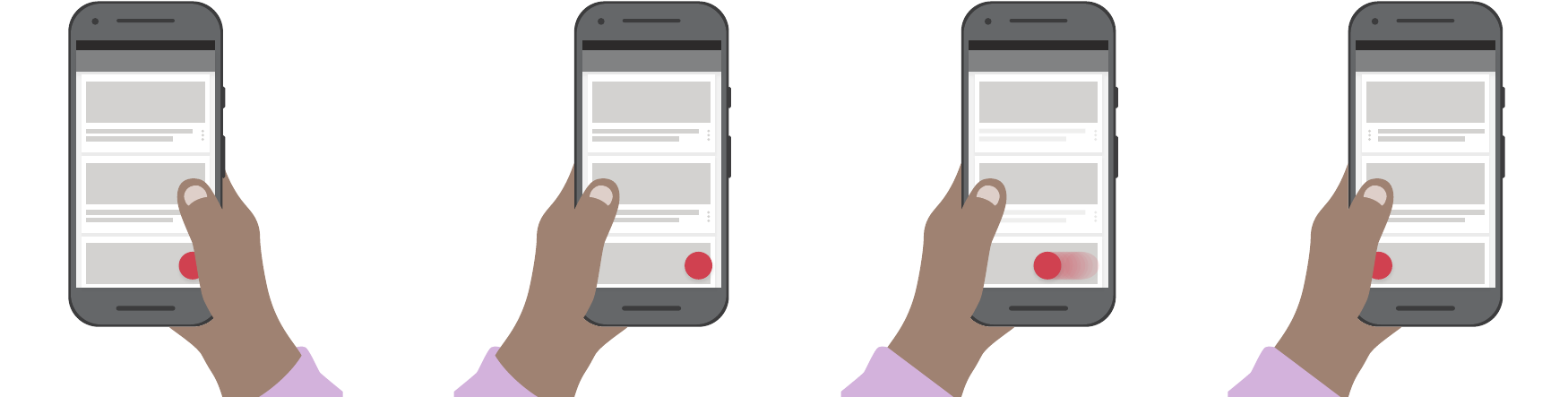}
\captionof{figure}{Mobile User Interface (UI) adapts to the handedness of the user and provides appropriate feedback to inform the user.
\label{fig:feature-graphic}}
\end{strip}

\begin{abstract}
Taller and sleeker smartphone devices are becoming the new norm. More screen space and very responsive touchscreens have made for enjoyable experiences available to us at all times. However, after years of interacting with smaller, portable devices, we still try to use these large smartphones on the go, and do not want to change how, where, and when we interact with them. The older devices were easier to use with one hand, when mobile. Now, with bigger devices, users have trouble accessing all parts of the screen with one hand. We need to recognize the limitations in usability due to these large screens. We must start designing user interfaces that are more conducive to one hand usage, which is the preferred way of interacting with the phone. This paper introduces Adaptive App Design, a design methodology that promotes dynamic and adaptive interfaces for one handed usage. We present a novel method of recognizing which hand the user is interacting with and suggest how to design friendlier interfaces for them by presenting a set of design guidelines for this methodology.
\end{abstract}
\category{H.5.2.}{Information Interfaces and Presentation}{User Interfaces -\textit{ graphical user interfaces}}

\section{Introduction}

The fashion industry is embracing and promoting the concept of "Cell Phone Pockets" that can fit 5+ inch mobile devices in pants and workout clothes. While this makes the smartphone easier to carry around, we must recognize that the large screen still decreases the degree of mobility these devices are meant to offer. People tend to use their phone in different contexts where variables in the environment influence how they hold the phone and interact with the applications. For example, people try to use their phone while walking, driving or eating which results in varying levels in attention and availability of fingers. When coupled with the large screen, usability becomes an important concern.

The touchscreen is the primary form of input on a smartphone. So the interaction model of a typical phone app relies on single/multi finger gestures as actions that objects in the app respond to \cite{nieters}. This includes gestures such as tap, double tap, swipe, pinch open or closed, double-finger tap, etc. and they can be performed on different elements positioned in different parts of the screen. Since we have to hold the phone with 1 or 2 hands whilst using it, we automatically restrict the number of fingers that are free to interact with the screen. This in turn restricts the number and type of gestures, and depending on how we hold the phone, the area of the screen that is reachable. For example, one way users hold their phone is with one hand where the thumb interacts with the screen while the 4 fingers grip the device. This limits the interaction model to one finger gestures. The length of the thumb further restricts the area reachable on the screen depending on the size of the phone \cite{bergstrom2014modeling}. To reach objects on the far side of the screen, the user has to resort to using both hands, changing their grip, or end up abandoning the task. This interaction model would be very different from one where the user places the phone on a surface to interact. The number of possible gestures increases by several factors and the entire screen is now reachable.

User Experience (UX) Designers are motivated to take advantage of the large and responsive multi-touch screens to provide rich content and functions that were previously only available on computers. However, currently they have a hard time designing a single UI that is both usable and enjoyable with these diverse interaction models \cite{karlson2008understanding}. Designers are forced to find a balance between usability and functionality, often compromising on usability. 

The concept of limited "reachability" in one-hand mode is recognized as an issue in the industry and is currently being addressed by solutions provided by the operating system or the hardware manufacturer rather than by individual apps. In the iPhone 6 and higher models for example, users can double tap the home button (or swipe down in iPhone X) to pull the entire screen down towards the bottom, allowing users to better access the top half of the screen. Some Android phones allow the user to swipe left or right along the action buttons at the bottom of the phone to shrink the interface and move it to the left or right bottom corner of the screen respectively. Some other proposed solutions include ThumbSpace \cite{karlson2007thumbspace} and The Fat Thumb \cite{boring2012fat}. ThumbSpace allows users to drag their thumb across a small and reachable portion of the screen to create a rectangle onto which the UI is projected and can be interacted with as usual. The Fat Thumb on the other hand proposes a solution for the lack of multi-finger gestures by using the rotation and pressure of the thumb to create new one finger gestures. Fat Thumb however still presents the problem of reachability.

The above solutions are all high level and generalized. These solutions are presented at the OS level and are agnostic of the application being used. Since apps vary widely in function and design, they are not always very effective in addressing reachability issues. Further, they limit the size of the interface, which is counter-intuitive to having the large screen. This wastes valuable real estate and decreases the size of touch targets. These are also manual modes that require the user's intervention each time an area is unreachable. All these factors take away from having enjoyable experiences.

Our solution gives individual app makers the power and flexibility to design reachable interfaces for different grips. It helps pinpoint the specific interaction model. Designers can then make conditional interfaces that provide different levels of functionality and different layouts in UIs (user interfaces) when for example, the phone is being held with one hand and used with a thumb, versus when the phone is on a surface and is being used with the index finger. This is analogous to responsive web design \cite{knight2011responsive}, where websites modify the layout of the pages and the type \& number of functions available to users based on the device they are using. For example, a document editing website may provide a complete set of text formatting capabilities when used on a computer while only providing the basic formatting functions when used on a mobile device. The UI components may be resized depending on the input mode (mouse vs touchscreen). Objects may also be rearranged based on screen orientation (landscape vs portrait). Therefore, the UI is conditional on the type of device being used and essentially the size of the screen.

In this paper, we introduce Adaptive App Design, a new design methodology that will lay the foundation for grip-friendly mobile interfaces. Using this method, similar to responsive design, apps can modify the layout of the screens and the type \& number of functions available to the user based on their grip on the device. We present a novel method of detecting which hand is interacting with the phone. By maintaining this data as a flag on the device, apps can then retrieve it to identify the interaction model. App makers can then design adaptive interfaces for each model. The UI is conditional to the grip and subsequently the reachable area of the screen. We present ground breaking findings from a primary usability study of this methodology and foundational design guidelines that app developers can use as a baseline for designing adaptive interfaces.

\section{Related Work}
\cite{holz2011understanding, karlson2005applens,weberg2001piece,wobbrock2008performance} are only few of many studies over several years stating that current interface designs on mobile devices are not suitable for one handed usage. Studies such as \cite{bergstrom2014modeling,lee2011fit} determine the functional area of the thumb, which quantifies the struggles that users face when operating these ill-designed interfaces.

Researchers have tried to identify which hand is holding the phone with the help of additional sensors or equipment. \cite{cheng2013igrasp,harrison1998squeeze,kim2006hand,taylor2008bar,wimmer2009handsense} used capacitive sensors along the sides of the phone to detect grip while \cite{taylor2009graspables} used additional accelerometers. \cite{mohd201428} on the other hand used the position of the fingers on the back of the phone to determine the finger interacting with the phone. This method also used additional capacitive sensors.

Our work aligns with \cite{goel2012gripsense,guo2016recognizing, lochtefeld2015detecting}, where they attempt to differentiate grips without having to use additional sensors. This allows the solution to be economically scalable since we do not have to augment the device with additional hardware and hence, we predict, would be more attractive for commercial companies to implement. However, all three teams collected data from a limited number of participants, and the differentiation of grips could be done only after the user interacted with the system for several steps. More importantly, all 3 solutions used machine learning kits to solve the problem. They have extracted features from data collected by the accelerometer, gyroscope and the touchscreen from tap and swipe gestures and fed them into a classifier. Our solution does not require machine learning, does not exert any additional load in terms of network access or additional computing modules, and has been tested with both left and right handed people. The computation required for it is very low and it can be easily implemented into the existing gesture classes (GestureDetectorCompat, UISwipeGestureRecognizer) in the APIs of Android and iOS platforms, allowing app developers to very easily adopt the system. While \cite{goel2012gripsense,guo2016recognizing,lochtefeld2015detecting} all heavily rely on swipe gestures for higher accuracy, they have collected data only in lab settings, and have not addressed the real world use cases in which the system will be activated. In the following sections, we present these use cases as well as present the design implications of having such a detection system in day to day applications.

\section{User Research}
As groundwork, we conducted qualitative and quantitative research to understand how users interact with their phone in non-lab settings. It consisted of 1) 20 hours of ethnographic research on a college campus by observing users interacting with their phones in settings such as public spaces (building lobbies, elevators, buses), offices, classrooms and while performing different tasks such as eating, walking, using a computer, etc. 2) a survey with 60 respondents to record self-reported data on their own mobile usage patterns 3) 11 in-person semi-structured interviews to understand and record specific problems and needs users face. In addition, we conducted an extensive study of design practices and guidelines followed by app designers in the industry to bridge the gap between user needs, design practices and available technology. The qualitative data from the ethnography and interviews was then coded, affinity mapped and analyzed alongside the ordinal data collected from the survey. The insights from these studies are summarized and discussed below.

\subsection{Handedness}

Although 90\% of the survey respondents were right-handed, only 70\% reported that they use their phone with their right hand. The rest use their phone with their non-dominant hand since they tend to interact with their phone during other passive activities such as eating or drinking. Interviews with users revealed that once they use their phone with their left hand, they place the phone towards the left of their computer on their desk, or in their left pocket or similar positions which further motivates them to use their phone with their left hand. This number is also not static since users reported not to be exclusively left or right handed when using their phones.

\subsection{Reachability}

\begin{figure}
\centering
  \includegraphics[width=1\columnwidth]{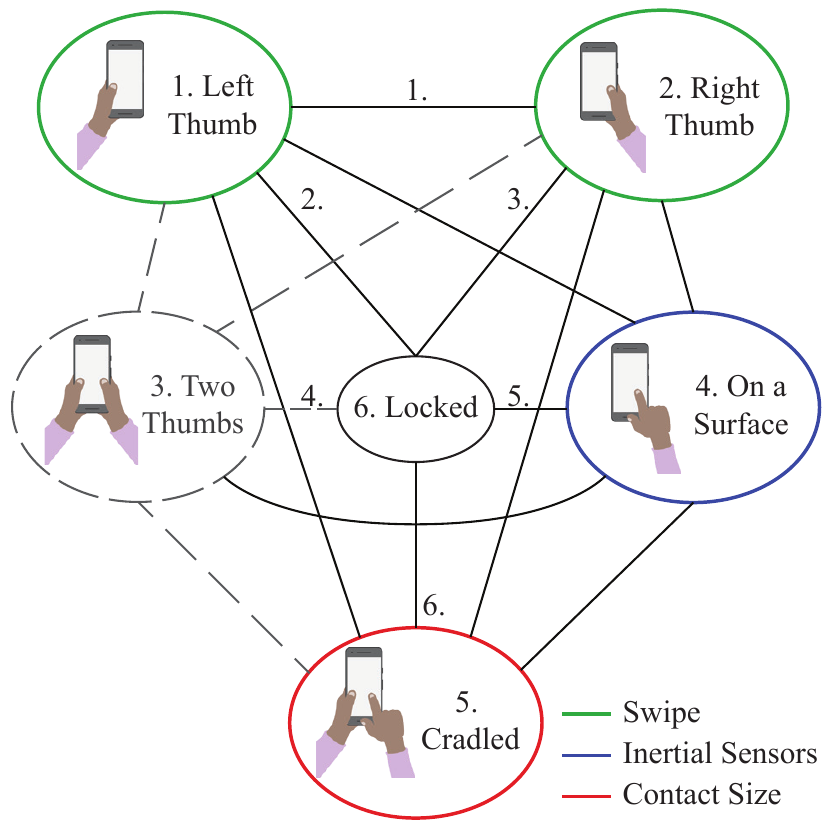}
  \caption{State Diagram of the various grips: Grips 1 \& 2 denote one-handed use with left and right thumb respectively (identified using swipe gestures), Grip 3 is when user interacts with both thumbs, Grip 4 is when the phone is placed on a surface (identified using inertial sensors), Grip 5 is when the phone is held in one hand and interacted with the other hand's index (identified by contact size) and Grip 6 is the locked state of the device. Transitions 1, 2, 3, 4, 5 denote unlocking the phone. }~\label{fig:states}
\end{figure}

The 5 common grips identified from the ethnography and interviews are as shown in Figure~\ref{fig:states}: 1. Held with the left hand and interacting with the left thumb, 2. Held with the right hand and interacting with the right thumb, 3. Held at the bottom with both hands and interacting with both thumbs, 4. Placed on a surface and interacting with an index finger, 5. Held with one hand (cradled) and interacting with the index of the other. Hoober \cite{hoober} identifies that the 6th mode of interacting with the device is in landscape mode. 68\% of the survey participants reported that they most often used their phone using grip 1 and 2 while 30\% reported that they use their phone in state 3 and less than 10\% used their phone in state 5. Each of these grips offer a varying level of reachability. Mapping the range of motion of the fingers shows that the area reachable changes with each of these grips. App designers refer to these as basis for organizing information and components on the screen. Many design for the general or the majority case which is right hand thumb usage, while others try to find a sweet spot on the screen that works for all the 5 modes \cite{smashing}. Not all apps can find a way to make this work and provide maximum functionality at the same time. Which is why we found that users must still reposition the phone in their hands to perform some tasks.
\begin{figure}
\centering
  \includegraphics[width=1\columnwidth]{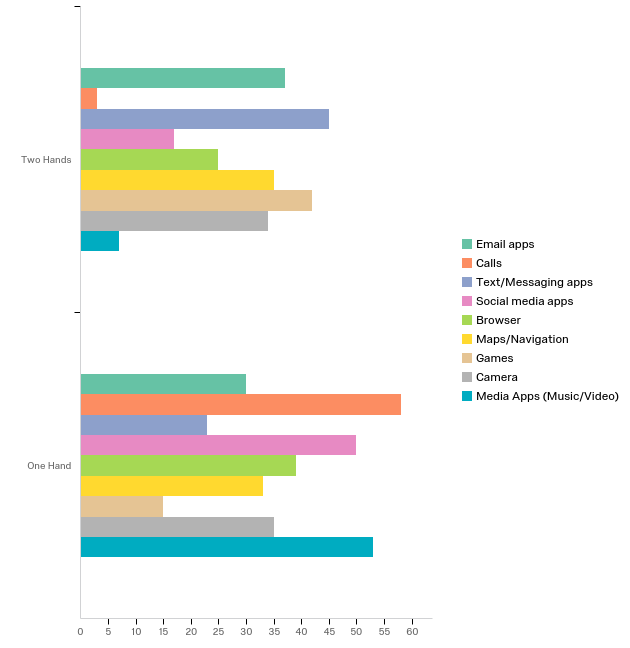}
  \caption{Bar chart denoting the how many hands they use to interact with common types of apps}~\label{fig:taskgraph}
\end{figure}
\subsection{Task/Function}
Participants in the survey reported that they tend to modify their grip based on the app they are using and the level of attention they dedicate to it. The bar chart in Figure~\ref{fig:taskgraph} quantifies how many hands people use to interact with certain types of apps. This data was further augmented by the interviews. For example, users are more likely to use a text messaging application by holding the phone with two hands and using their thumbs. This is because of the frequency of tapping that is required, and the speed at which they interact with the screen. Despite the availability of one-handed keyboards, users still prefer to use both hands since the target size of one-handed keyboards reduces. Users are less likely to perform this activity if both hands are not free to handle the phone. The same applies to gaming apps. On the other hand, users are more likely to use social media apps with one hand and thumb to scroll through information. Users tend to pay less attention during this activity and tend to use these apps more when on the go.
\subsection{Environment}
We observed that users tend to change the way they interact with their device based on the context or the environment. When driving, they interact with the device with less frequent one-off gestures with their index finger. Similarly, during passive activities like drinking and eating they might focus their attention to the phone but interact with their device with their thumb, or if the device is placed on a surface, with their index. In social contexts, when the user is in the presence of other people, they are more inclined to use their phone with one hand in short sessions, paying little attention to the device. By probing to uncover more insights in the interviews, we also learned that users tend to hold their phone in one hand while walking but specifically tried to look at their phone less and in bursts since their visual focus is hindered. Even the cold weather influences how many hands they are willing to expose to use their phone. 

\subsection{Discussion}

The above factors put together influence the interaction model of the user with the device and subsequently the app. Based on these findings we propose that the ideal mobile interface design should be defined as:
\begin{equation*}
\textnormal{Mobile User Interface} = f(task, context, grip)
\end{equation*}
i.e., good interfaces could be designed if the phone understands the task, context and grip. The task is the action that has to be performed by the user in the app, which app designers are aware of. The context encompasses the environmental variables such as the social presence around the user and weather, as well as the passive or concurrent activity the user is engaged in when interacting with the phone. This could be walking, eating, cooking, etc. Understanding this context proves to be more of a challenge. It can be inferred in some scenarios like walking, driving, etc. through signal analysis of the inertial sensors of the phone. More work is being done in the space of mobile and ubiquitous computing and activity tracking to use existing sensors on the phone \cite{abowd1997cyberguide, kwapisz2011activity, lu2017towards}, as well as using other wearable and mobile devices with audio, haptic and inertial sensors \cite{bao2004activity, chen2012sensor, maurer2006activity} to become aware of context and activities such as eating, typing and conversing. While the context influences the grip on the device it is not a direct causality. The grip is often a user preference. We've observed this in the case of typing. Similarly, when using video player apps, even though the users focus is on the app and the context may be conducive to two handed usage such as in waiting rooms or airports, users reported that they preferred using one hand. Therefore grip must be considered as an independent variable when designing the interface. 

Less work is being done to study grip. Therefore, we focused our research on this aspect. We started with the observation that when people use their phone with their left or right thumb, they tend to tilt their swipe, trace or fling gestures towards the left or right respectively. Therefore, if we can sufficiently differentiate between these swipe gestures, we would be able to determine which thumb is being used to interact. We then went on to brainstorm other possibly generalizable characteristics of touch gestures (tap, drag, scroll, etc.) and input from sensors including fingerprint and inertial sensors that already exist on modern mobile devices that may be able to identify grips. This lead us to the state diagram shown in Figure~\ref{fig:states}.

Here, each node represents a grip or holding pattern. The edge represents the transition from one grip state to another. For app developers to be able to make adaptive interfaces for each state, they must be aware of when a user changes from one grip to another. Therefore, identifying those transitions is key. From the ethnography and interviews, we uncovered that users tend to maintain the same type of grip for a single session of use, where a session is the period from the point a user unlocks their phone to the point of locking the phone or putting it away. Also, users tend to perform similar type of tasks that require a similar type of grip per session especially when it is dictated by the context. We observed that the two main tasks that warrant a change in grip in the same session was typing or playing a game. This was a causation of the task rather than the grip or context. Therefore identifying the unlocking transitions is important which prompted us to look into using the various unlocking mechanisms. Further, major reachability issues occur in states 1 and 2 since the reachable area is severely limited. This coupled with the fact that they are the most commonly used grips made transitions 2 and 3 a priority for us. We focus this paper on identifying these transitions and our next steps would be to evaluate the accuracy of the inertial sensors and contact size as a method of detection.

On phones that use biometric unlocking, when the user is asked to register their fingerprints on setup, the phone can register left and right digits and subsequently use that data to determine which hand is being used to interact with the phone. This holds true for both front and back fingerprint sensors. For devices that don't use fingerprint sensors, we can use the swipe-up to unlock screen or the scroll gesture. Because the thumb is hinged when opposed to the 4 fingers, it has limited lateral range of motion. This is reflected in the swipe and scroll gesture. These methods cover the common unlocking mechanisms including the facial recognition system on iPhone X where users are still presented with a swipe up to unlock screen.

While these help us identify the unlocking transitions, we also discovered that the swipe or scroll gesture is a very commonly used one. Since the phone is primarily used in portrait mode, most applications (social, media, email, etc.) are designed to scroll vertically to present content. In an interview, one user described her Instagram usage as "a meter long scroll". \cite{leo} states that on an average, smartphone users scroll 201 to 257 times a day. This number is specific to applications that present content on the Internet such as social media apps, websites, articles, news apps, etc. In addition to these there are other apps that also employ the linear list type presentation of content such as email, shopping etc. and even apps native to the phone such as the Settings app. The scrolling gesture is also used in the app drawers. Therefore, the scroll is the most common gesture alongside the tap gesture. Tracking these gestures would give us an accurate and continuous inference of which hand is interacting with the screen.

% Use a numbered list of references at the end of the article, ordered
% alphabetically by first author, and referenced by numbers in
% brackets~\cite{ethics, Klemmer:2002:WSC:503376.503378,
%   Mather:2000:MUT, Zellweger:2001:FAO:504216.504224}. For papers from
% conference proceedings, include the title of the paper and an
% abbreviated name of the conference (e.g., for Interact 2003
% proceedings, use \textit{Proc. Interact 2003}). Do not include the
% location of the conference or the exact date; do include the page
% numbers if available. See the examples of citations at the end of this
% document. Within this template file, use the \texttt{References} style
% for the text of your citation.

% Your references should be published materials accessible to the
% public.  Internal technical reports may be cited only if they are
% easily accessible (i.e., you provide the address for obtaining the
% report within your citation) and may be obtained by any reader for a
% nominal fee.  Proprietary information may not be cited. Private
% communications should be acknowledged in the main text, not referenced
% (e.g., ``[Robertson, personal communication]'').

\section{Swipe Detection}

We started by collecting real time swipe gesture data by sending out 2 Android apps to 7 participants (1 left handed) that would collect \textit{x, y} coordinates and timestamps of the points on a swipe gesture when being used. They were installed on the participants own device and used in their daily contexts and during their regular activities. The first app mimicked the swipe up to unlock application. It would pop up upon unlocking where the user would have to swipe the screen up and then click one of 4 buttons to indicate if the gesture was performed by the left index, right index, left thumb or right thumb. This way the app was used on an average of 16 times a day by each user for 7 days resulting in ~800 data tuples. The second app was a scrolling app which consisted of a large block of text in the form of a story. Users were asked to first indicate through radio buttons if they were using their left index, right index, left thumb or right thumb and then asked to scroll through the text. If they were to switch fingers, they could select the button for the corresponding finger they were going to use. This resulted in ~600 data points. All the data points were collected on a real-time database and analyzed separately.

Upon visualizing the data collected, we observed that the left thumb and right thumb swipes are very distinct in curve but not position. It appears users with longer thumbs tend to swipe on the opposite half of the screen from their hand whereas users with shorter thumbs swipe on the portion of the screen closer to them. The slopes of the curves however, are distinct. So instead of approaching this as a machine learning classification problem we looked at the data as geometric curves. If we could find templates that represented a left and right swipe we could compare the slope of subsequent swipes with these templates.

We used the curve fitting method called Polynomial Regression \cite{heiberger2009polynomial}, commonly used by statisticians to determine the relationship between an independent variable \textit{x} and a dependent variable \textit{y}. Using this method, a polynomial of degree \textit{m} is defined that would most closely match recorded values with the values predicted by the polynomial. This matching is done by reducing the variance calculated from the ordinary least-squares method. The resulting polynomial will be of the form:
\begin{equation*}
y = \beta_1 + \beta_2x + \beta_3x^2 + ... + \beta_{m+1}x^m
\end{equation*}
where \begin{equation*}
\beta_1,\beta_2,..\beta_{m+1} 
\end{equation*}
are the coefficients that will act as identifying properties of the curve. Therefore, for our purposes, the range of values for these coefficients should be sufficiently distinguishable for the left and right swipe curves to accurately determine which thumb was used. 

The modification we made for our purposes was to have \textit{y} as the independent variable and \textit{x} as the dependent variable to match the orientation and coordinate system of the phone. The polynomial would take the form:
\begin{equation*}
x = \alpha_1 + \alpha_2y + \alpha_3y^2 + ... + \alpha_{m+1}y^m
\end{equation*}
We used the Python package NumPy's polyfit() \cite{scipy} method. Running the method for degrees 1, 2, 3 and 4 on the data gave us a range of \textit{r}, root mean square error (variance) values and by comparing them, we found that \textit{m = 2} gave \textit{r} values consistently greater than 0.9, giving us an almost perfect match between the estimated curve and the actual curve. 
\begin{figure}
\centering
  \includegraphics[width=1\columnwidth]{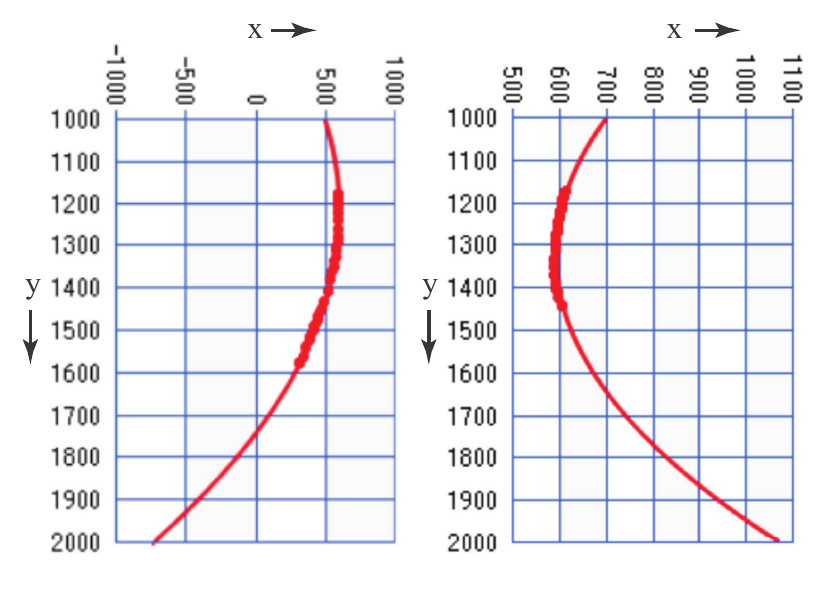}
  \caption{Visualization of the polynomials obtained from Quadratic Regression of a sample left and right swipe.}~\label{fig:swipes}
\end{figure}
Therefore, the polynomial defining the swipe gestures is of the form:
\begin{equation*}
x = A + By + Cy^2
\end{equation*}
where \textit{A, B} and \textit{C} are the features of the curve. This defines the swipe curve as a quadratic polynomial or a parabola which opens to the left or right for left or right swipes respectively as shown by Figure~\ref{fig:swipes}. Parabolas that open to the left would have the value \textit{C} as negative whereas parabolas opening to the right have the value \textit{C} as positive. This was consistent for 99.53\% of the data points collected for left and right thumb swipes.

Therefore, all we need to do to differentiate between the left and right thumb swipes would be to implement a Quadratic Regression calculator, an \textit{O(n)} algorithm where \textit{n} is the number of points collected by the swipe event on the device. This requires very little computation as \textit{n} will always be finite for gestures performed in real world scenarios. The device is already collecting these points by monitoring the touch screen for activity and therefore no additional battery consumption would take place. Also, no data needs to be transmitted across a network since all the calculation would be done on the device in the same class that registers the motion event gesture. This we predict, would be a big motivation for mobile phone manufacturers to implement this system.

\section{Evaluation}

To test the system in real world scenarios, we decided to evaluate it from the lens of multiple stakeholders. This gave us an idea of not only how feasible and useful the solution is but also if the stakeholders would adopt it and how difficult that process would be.
\subsection{Technical Evaluation}
As mentioned earlier, the system consists of a low complexity \textit{O(n)} snippet that does not require any additional data from the ones already being collected in the GestureDetectorCompat, UISwipeGestureRecognizer classes in Android and iOS respectively. The snippet can be implemented as a method in these classes that runs in the background when a swipe gesture occurs and maintain a flag/state to indicate left and right 
thumb usage. Apps can then either poll this state, or receive a broadcast from the OS and dynamically change the layout. Developers can then modify the layout files to rearrange the interface without having to change the inherent functionality of the application and therefore will have little overhead. 
To test this aspect along with the accuracy of the system, we implemented an Android app consisting of a long list of items similar to the Settings app, where each item consists of text and controls. We implemented the quadratic regression as a method that was run asynchronously after a swipe occurred using the least squares method implementation \cite{keisan} and broadcast an event when the state of the grip changes. On receiving the event, the arrangement of items in the list was reversed from left to right. We tested the application with 12 users (2 left handed) with thumb lengths ranging from 2.4 to 2.9 inches and the system could accurately distinguish 194/196 swipe gestures. The algorithm worked well even when the participants used their non-dominant hand and when swiping at different points of the screen and different angles. We noticed that this method worked irrespective of whether the participants griped the bottom of the phone or higher/towards the center of the phone. There was also no lag in response from the device.

\subsection{Design Evaluation}
This feature opens new degrees of freedom to app designers. It provides insight that better informs the interaction model between the user, the device, and subsequently the app. To understand how designers would be able to leverage the concept of Adaptive App Design, we conducted a design jam with 6 mobile app designers (1 left handed) who are familiar with the iOS and Material design guidelines \cite{IOS,Material}. Each designer had experience working on apps with varying levels of functionality and constraint in screen space. The insights were therefore fairly generalizable to all apps. It was a semi-structured session where we first explained the technology and what it would output. Second, we presented screens from common everyday apps as examples to consider. We then presented design prompts - modifications on these screens - that would work with this feature. The participants were then asked to critique, modify, re-design or discard those designs in a focus group/design brainstorming style discussion. The resulting insights, along with insights from usability testing (below section) were coded and affinity mapped. The trends and patterns noted were then consolidated into the design guidelines outlined later in this paper.
\begin{figure}
\centering
  \includegraphics[width=1\columnwidth]{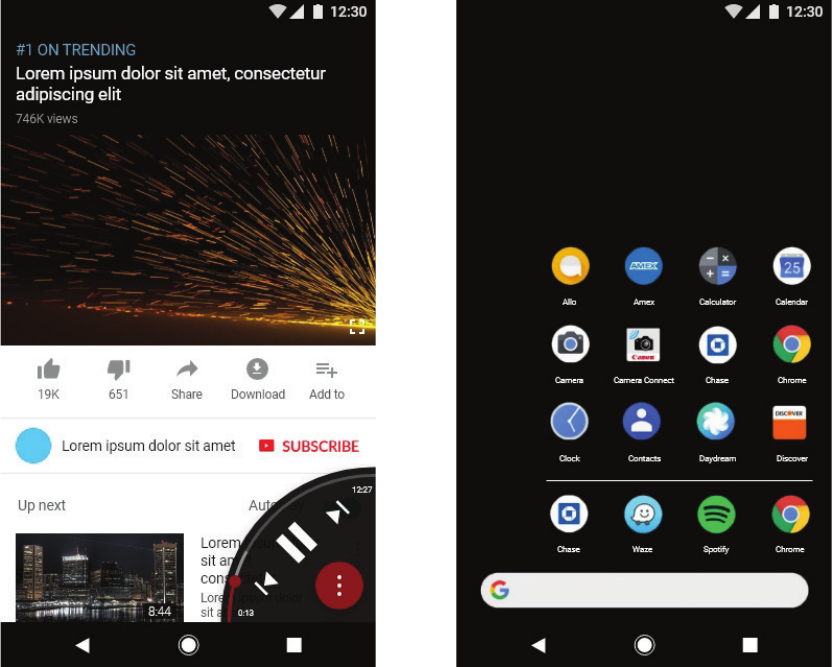}
  \caption{Screens from prototype used for user testing. (left) Video Player with controls designed for right thumb usage. (right) App Drawer with icon grid designed for right thumb usage.}~\label{fig:designs}
\end{figure}
\subsection{Usability Evaluation}
We took inspiration from the design jam and created a prototype of a video player application in the Android platform in which the interface would rearrange itself based on which hand was being used to interact. We chose a media app since our initial research indicated that people prefer using such apps with one hand. Each screen was designed to be suited for thumb usage by concentrating UI components to the reachable corner of the screen. Figure~\ref{fig:designs} showcases two screens designed for right thumb use. Elements that are concentrated to the right move to the left for left thumb use. We used existing components such as buttons, menus etc. to match current Android design standards (Material) \cite{Material}. When the screen rearranged, we made sure to use smooth transitions defined by the Material design guidelines to maintain consistency. We used different types of transitions such as move, fade, shrink, etc. for different elements on different screens to see how users responded to each. In addition, we mimicked the android operating system UI and the app drawer which we modified to be more reachable. We also applied the same principles of rearranging elements on the home screen and app drawer based on the hand.

Users were then asked to perform tasks with the application such as search for a video, change settings, open a trending video, etc. in the following order:
\begin{enumerate}
\item Interacting with the app by holding the device in their dominant hand. This was to obtain their unbiased opinion of the one-hand mode design of the app.
\item Interacting with the app by holding the device in their non-dominant hand. This gave us insight on their level of comfort using the rearranged interface with the hand they use less frequently.
\item Switching hands while interacting. This was to understand how they feel about the screen rearranging, the response times, as well as to see if they could form a mental model of how this feature works.
\end{enumerate}
We conducted this session with 15 (3 left handed) participants in lab settings. The participants were asked to think aloud while interacting with the device. The post-test evaluation involved understanding three aspects: desirability, attractiveness and usability.

To measure desirability, we used a subset of the Microsoft reaction cards deck \cite{benedek2002measuring} and asked participants to pick 5 out of 25 (both negative and positive) descriptive words that would help articulate their thoughts about the feature. They were then asked to explain why they chose those words forcing them to expand on their emotions. By doing this task, users were given the complete flexibility to choose the qualities they associate with the feature instead of us limiting our evaluation to the degrees of certain qualities.

The participants were then asked to fill Likert scale questions which were modified versions of the AttrakDiff Lite questionnaire \cite{hassenzahl2003attrakdiff} and the System Usability Scale \cite{brooke1996sus} to measure the ease of use as well as the perceived control users feel they have over the system. When put together the three post-test evaluations gave us a preliminary idea of how willing users are to adopt a system like this, how seamless the transition to this system would be and whether the adoption of this feature into everyday activities would be intuitive and not require much thought. Since we were introducing a novel feature, the goal of this study was to understand initial reactions to the concept of adaptive interfaces and iterate to see how we may improve usability with more refined guidelines.

\section{Results and Design Guidelines}
The overall reception to this feature was positive with 93\% of responses stating that they found the system appealing and they would like to adopt it into their devices. The most common words used to describe their experience with the system were effortless, useful and accessible. When asked to elaborate, users explained that they didn't have to pay much attention to the fact that the switch is happening since it only happens when their attention is interrupted by them switching hands. They also do not have to exert effort cognitively or physically to make the components more reachable. Most importantly, all the participants pointed out that they now didn't have to use their other hand or juggle the phone to interact with the screen since everything they wanted to do was within reachable distance from their thumb. The Pragmatic Quality (PQ) score of 1.71 from the Attrakdiff questionnaire backed this finding. This suggests that for these tasks in current static designs, using both hands was more of a coping mechanism. However, we intend to do further research to see if people will modify the way they use their phone given this new adaptive quality of interfaces.

33\% of the participants were initially thrown by the interface changing since they weren't expecting it to and the remaining participants were surprised but experimented more to try to understand the trigger of the change. However, all the participants responded that they subsequently saw it coming. They were also able to understand the model easily after a few interactions. Further 93\% of the participants indicated that it would be easy to learn the system quickly. This was also indicated by the SUS learn-ability score of 3.83/4. One user compared it to the first time she encountered the Portrait to Landscape orientation feature in phones and how after the first time it happened she knew to look for it.

Left handed participants were particularly receptive since they believed that most current UIs are designed for the right-handed user. Because of this, they tend to hold the phone in their right hand but have to resort to using their more stable left hand to interact with the screen. Now with this feature, they can use the device with their left hand alone.

Overall this study pointed out that the feature would make the phone more usable on the go and would fit in seamlessly into people's daily usage without disrupting their activities more than having to update their phone and see minimally redesigned applications. The attractiveness (ATT) score was 1.75 and was higher than the hedonic scores. This was expected as this feature is meant to run in the background and should not need or call for the user's attention. However, the high ATT score suggests that adaptive interfaces are a real need.

Designers were encouraged by the possibilities this opens to making screens less restrictive in space and size. By understanding the grip of the users, they were not just receptive to the idea of knowing what area of the screen is reachable but also pointed out that they could now map the area of the screen that is not hidden by the thumb or thumb movement which can help with targeting content. They pointed out that making screens more usable with one hand and on the go could potentially increase face time.

By generalizing the insights gained from the design jam and user testing we recommend the following non-exhaustive list of formative guidelines for Adaptive App Design for one handed usage. These guidelines only pertain to the grip and does not talk about the context of use or the task being performed and we encourage designers to consider these aspects in addition to the below guidelines to create good mobile user interfaces.

\begin{enumerate}
\item Minimize change blindness by providing feedback. Indicate what has changed on the interface and how it has changed. This is in line with Neilson \& Molich's guideline to maintain visibility of system status \cite{nielsen1990heuristic}. Use transitions and animations as shown in Figure~\ref{fig:feature-graphic} to indicate to the user that there are changes being made to the interface.
\item Provide timely feedback by avoiding delays in rearranging the interface to make sure that the response from the system feels real time or feels like a direct result of an action performed.
\item Use different responses and animations from the ones native to the functionality of the applications to indicate that the response/transitions correspond to the change in handedness rather than a function on the screen. 
\item To maintain spatial and hierarchical relationships between elements on the screen, ensure that the elements that change are identical in both left hand state and right hand state.
\item Group elements that would get rearranged by using layers/shadows or other demarcations to indicate to the user that those elements will move. Maintain separate layers for elements that don't get rearranged. For example, in Figure~\ref{fig:feature-graphic}, the floating action button at the bottom of the screen appears to be on a separate layer from the other content allowing it to move freely across the screen.
\item Design new components rather than modifying existing components that work well for both left and right handed UIs. Keep in mind that it is not the same as mirroring all aspects of the screen as is the case when designing for right to left (RTL) reading languages. Users will still read information from the left to right but can interact from either the left or right side of the screen.
\item Order components based on frequency of use and importance having the most important or most frequent actions more accessible to the thumb.
\item Do not rearrange temporal aspects of the interface. Change in time is mapped from left to right in LTR languages and relies on prior knowledge. Therefore, this aspect should be maintained through the interface. For example, the back button should continue to point to the left. However, the position of the back button may be changed.
\item Do not rearrange elements that are not immediately visible. Components that are temporarily hidden like for example menus that appear on swipe left or right should remain unchanged as there is no visual indication that it could be rearranged. Gestures that directly map to their appearance on screen should also be maintained as in the case of the swipe left and swipe right. However, trigger buttons may be repositioned on the screen.
\item Do not change the direction of elements that scroll horizontally. Content in these containers must be revealed in the same direction as it is read.
\item Since the main aspect of the interface that is changing is the position of elements on the screen use translation/move animations extensively to indicate what is moving and to where. Use principles of easing, offset and parallax to maintain hierarchy of elements.
\item Avoid placing intractable elements in the areas that are not reachable. Make them, at minimum, partially overlap with the reachable areas by increasing the size of elements or placing bigger elements farther. If not, introduce redundancy in functions through collapsible menus placed closer to the thumb.
\end{enumerate}

\section{Conclusion and Future Work}
In this paper, we presented Adaptive App Design - a novel methodology for designing dynamic interfaces for changing grips. We presented the factors influencing interaction models between users and mobile devices and identified that good mobile interface design would be a function of the task, context and grip on the device. We presented a method to detect the grip that would work with this design methodology and outlined preliminary guidelines to design adaptive interfaces for left and right hand usage. We believe that the above solution and guidelines would make interfaces more usable overall. However, this is still a new concept and we hope that more research will be focused in this area. We have yet to evaluate the accuracy of detecting the other common grips in holding the device that offer more flexibility in the interaction model and our next steps would focus on this area. We plan on conducting more user research to see how users respond to adaptive interfaces for these additional grips. But the ideal adaptive interface would still consider the context and task as well as the various other grips. The next steps would be to consider technology and innovation that would allow the phone or other devices to be fully aware of these three parameters following which it is important to bridge the gap between the technology and design. We would like to further augment these design guidelines with information to design apps that are also aware of the context or activity the user is engaged in while using the phone. Our ultimate goal is to design the ideal mobile interface that is aware of task, context and grip.

% Balancing columns in a ref list is a bit of a pain because you
% either use a hack like flushend or balance, or manually insert
% a column break.  http://www.tex.ac.uk/cgi-bin/texfaq2html?label=balance
% multicols doesn't work because we're already in two-column mode,
% and flushend isn't awesome, so I choose balance.  See this
% for more info: http://cs.brown.edu/system/software/latex/doc/balance.pdf
%
% Note that in a perfect world balance wants to be in the first
% column of the last page.
%
% If balance doesn't work for you, you can remove that and
% hard-code a column break into the bbl file right before you
% submit:
%
% http://stackoverflow.com/questions/2149854/how-to-manually-equalize-columns-
% in-an-ieee-paper-if-using-bibtex
%
% Or, just remove \balance and give up on balancing the last page.
%

\balance{}

% BALANCE COLUMNS
\balance{}

% REFERENCES FORMAT
% References must be the same font size as other body text.
\bibliographystyle{SIGCHI-Reference-Format}
\bibliography{references}

\end{document}